\documentclass{article}
\usepackage{graphicx}

\usepackage{jheppub}
\usepackage{amsmath}
\usepackage{epsfig,multicol,bbm}
\usepackage{url}

\usepackage{graphicx}
\usepackage{setspace}
\usepackage{amssymb}
\usepackage{listings}
\usepackage{epstopdf}
\usepackage{float}
\usepackage{ulem}
\usepackage{color}
\usepackage{amsmath}
\usepackage{ dsfont }
\usepackage{slashed}

\usepackage{amsmath}

\title{Focus Point Supersymmetry in Extended Gauge Mediation}

\date{\today}

\author[1]{Ran Ding}
\author[2,3]{Tianjun Li}
\author[4]{Florian Staub}
\author[1]{Bin Zhu}
\affiliation[1]{School of Physics, Nankai University, Tianjin 300071, P. R. China}
\affiliation[2]{State Key Laboratory of Theoretical Physics
and Kavli Institute for Theoretical Physics, China (KITPC)\\
Institute of Theoretical Physics, Chinese Academy of Sciences,
Beijing 100190, P. R. China}
\affiliation[3]{{School of Physical Electronics,
University of Electronic Science and Technology of China,
Chengdu 610054, P. R. China}}
\affiliation[4]{Bethe Center for Theoretical Physics \& Physikalisches Institut der
 Universit\"at Bonn, \\
Nu{\ss}allee 12, 53115 Bonn, Germany}

\emailAdd{dingran@mail.nankai.edu.cn}
\emailAdd{tli@itp.ac.cn}
\emailAdd{fnstaub@th.physik.uni-bonn.de}
\emailAdd{zhubin@mail.nankai.edu.cn}

\abstract{We propose a small extension of the minimal gauge mediation through the combination
of extended gauge mediation and conformal sequestering. We show that the focus point
supersymmetry can be realized naturally, and the fine tuning is significantly reduced
compared to the minimal gauge mediation and extended gauge mediation without focus point.
The Higgs boson mass is around 125 GeV,  the gauginos remain light, and the gluino is
likely to be detected at the next run of the LHC. However,
 the multi-TeV squarks is out of the reach of the LHC.
The numerical calculation for fine-tuning shows that this model remains natural.}

\begin{document}

\maketitle

\section{Introduction}
\label{sec:intro}

The discovery of the Standard Model (SM) like Higgs boson~\cite{Aad:2012tfa,Chatrchyan:2012ufa}
with $m_h\simeq125$~GeV has profound implications on naturalness for the
minimal supersymmetric standard model (MSSM). In the context of the MSSM there is a strict upper
bound on the light Higgs mass at tree level given by
$m_h\leq m_Z$ (see e.g. \cite{Carena:2002es,Djouadi:2005gj} and references therein). Thus,
the large radiative corrections, mainly from (s)tops, are necessary to lift $m_h$ to
the desired range of about 123--129~GeV. The dominant one-loop corrections can be approximated as \cite{Carena:1998wq}
\begin{align}
\label{mhtb}
 m_h^2=m_Z^2\cos^2 2\beta + \frac{3m_t^4}{4\pi^2 v^2} \left(\log\left(\frac{M_S^2}{m_t^2}\right)+\frac{X_t^2}{M_S^2}\left(1-\frac{X_t^2}{12 M_S^2}\right)\right) \, .
\end{align}
Here, $M_S=\sqrt{m_{\tilde{t}_1} m_{\tilde{t}_2}}$ is the supersymmetry (SUSY) scale defined as
the geometric mean of the two stop masses, $m_t$ is the running top quark mass, and $X_t$ parametrizes
the left-right mixing in the stop sector.
One widely used possibility to maximize these corrections is to consider a maximal mixing ($X_t \sim \sqrt{6} M_S$) in the stop sector while assuming only moderately large stop masses, see for instance \cite{Djouadi:2013lra,Carena:2013qia} and references therein. In that case and including two- and three-loop corrections \cite{Degrassi:2001yf,Brignole:2001jy,Brignole:2002bz,Dedes:2002dy,Dedes:2003km,Kant:2010tn} it is possible to explain the Higgs mass with stop mass around 1~TeV.
However, it has recently been pointed out that a maximal mixing in the stop sector can lead to a global minimum in the scalar potential at which charge and color are broken by vacuum expectations values (VEVs) of the stops  \cite{Camargo-Molina:2013sta,Blinov:2013fta,Chowdhury:2013dka}. The electroweak vacuum will only be metastable and could decay in a cosmological short time. Thus, one is tempted to choose the other possibility to enhance the radiative corrections by using heavier stop masses but keeping the left-right mixing small.
In order to accommodate for a Higgs mass in the desired range, stop masses $\geq 5$~TeV are needed in this scenario. Together with the lack of a signal of any new
physics at the LHC this raises uncomfortable issues with naturalness which is widely discussed in the literatures~\cite{Barbieri:1987fn,Anderson:1994dz,Cohen:1996vb,Ciafaloni:1996zh,Bhattacharyya:1996dw,Chankowski:1997zh,Barbieri:1998uv,Kane:1998im,Giusti:1998gz,BasteroGil:1999gu,Feng:1999mn,Romanino:1999ut,Feng:1999zg,Chacko:2005ra,Choi:2005hd,Nomura:2005qg,Kitano:2005wc,Nomura:2005rj,Lebedev:2005ge,Kitano:2006gv,Allanach:2006jc,Giudice:2006sn,Perelstein:2007nx,Allanach:2007qk,Cabrera:2008tj,Cassel:2009ps,Barbieri:2009ev,Horton:2009ed,Kobayashi:2009rn,Lodone:2010kt,Asano:2010ut,Strumia:2011dv,Cassel:2011tg,Sakurai:2011pt,Papucci:2011wy,Larsen:2012rq,Baer:2012uy,Espinosa:2012in}.

One can easily understand this issue from the tree-level condition of electroweak symmetry breaking (EWSB) which relates the Higgs soft-breaking masses, the $\mu$ parameter and $m_Z$. For $\tan\beta\geq 5$ the condition can be expressed by
\begin{align}
m_Z^2\approx -2(\mu^2+m_{H_u}^2[m_w])~.~\,
\end{align}
Often one does not take $m_{H_u}^2[m_w]$ but $m_{H_u}^2[\Lambda]$ as input. $\Lambda$ is the scale where SUSY is broken by some interactions with a hidden sector. The values for $m_{H_u}^2$ at $m_w$ and $\Lambda$ are connected by the renormalization group equations (RGEs). For the evaluation of $m_{H_u}^2$ the stop masses play an important role because of the size of the top Yukawa coupling $y_t$. One finds the relation
$m_{H_u}^2[m_w] = m_{H_u}^2[\Lambda] + \delta m_{H_u}^2$ with
\begin{align}
\delta m_{H_u}^2\sim -\frac{3y_t^2}{8\pi^2}(m_{Q_3}^2+m_{U_3}^2+A_t^2)\log\left(\frac{\Lambda^2}{m_{w}^2} \right)~.~\,
\label{eqn:deltaHu}
\end{align}
In this approximation, we have only considered the third generation Yukawa couplings but neglecting contributions
from gaugino masses.
The large contributions of the stop masses to the running of $m_{H_u}^2$ demand some fine-tuning of the fundamental parameters to get viable EWSB. To quantify this fine-tuning different measures have been introduced.  We are using throughout this work the one proposed by Barbieri-Giudice \cite{Barbieri:1987fn,Ellis:1986yg}
\begin{align}
\label{eq:FT}
\Delta_{BG} \equiv \max \{ \Delta_a\} \;\; \mbox{ where }\;\; \Delta_a \equiv \frac{\partial \log m_z^2}{\partial \log a} \, .
\end{align}
$a$ are the fundamental parameters in the theory. For the constrained MSSM (CMSSM)\cite{Ohta:1982wn} one takes $a\in \{ m_0^2,\; m_{1/2}^2,\; A_{0}^2,\;\mu^2,\;B\mu\}$. In this measure the overall fine-tuning of the MSSM in the context of squark masses above 5~TeV and small mixing is expected to be above $10^4$ \cite{Baer:2012mv}.

This large fine-tuning in the MSSM has triggered a lot of interests in models which already increase the tree-level Higgs mass by new contributions from F- or D-terms \cite{Ellwanger:2009dp,Ellwanger:2006rm,Ma:2011ea,Zhang:2008jm,Hirsch:2011hg,Bharucha:2013ela}. Especially in singlets extensions like the NMSSM  \cite{BasteroGil:2000bw,Dermisek:2005gg,Dermisek:2006py,Dermisek:2007yt,Ellwanger:2011mu},  GNMSSM \cite{Lee:2010gv, Lee:2011dya,Ross:2011xv,Ross:2012nr,Kaminska:2013mya} or DiracNMSSM~\cite{Lu:2013cta,Kaminska:2013new}, the fine-tuning is several orders smaller than in the MSSM. 

However, also in the MSSM exists parameter regions in which the fine-tuning becomes significantly smaller by one to two orders compared the general expectations. These are the focus point (FP) regions \cite{Feng:1999mn,Feng:1999zg,Feng:2000bp,Feng:2011aa,Feng:2012jfa}. In FP supersymmetry (SUSY), $m_{H_u}^2[m_w]$ is generated naturally and to a large extent insensitive to the variations of fundamental parameters at the scale $\Lambda$. Besides FP SUSY in the CMSSM, there are also investigations in other SUSY-breaking models including gauge-mediated supersymmetry breaking (GMSB)~\cite{Agashe:1999ct,Brummer:2013yya,Zheng:2013kwa}, models with large gaugino masses\cite{Abe:2007kf,Horton:2009ed,Younkin:2012ui,Yanagida:2013ah}, and hyperbolic branch SUSY\cite{Chan:1997bi,Akula:2011jx}. We are going to consider here SUSY breaking in the visible sector triggered by gauge interactions.

Already the minimal version of the gauge mediated supersymmetry breaking (GMSB)~\cite{Dine:1981gu,Dine:1981za,Dimopoulos:1981au,Nappi:1982hm,AlvarezGaume:1981wy,Dine:1993yw,Dine:1993qm,Dine:1994vc,Dine:1995ag} has the appealing features that it softens the flavor problem present in gravity mediated SUSY breaking scenarios \cite{Raby:1997pb}. On the other hand the minimal GMSB has became unattractive after the Higgs discovery since the $A$-parameters are only generated at the two loop level and usually negligible. Hence, even larger stop masses are needed than in the CMSSM with moderate $A_0$ to explain the Higgs mass \cite{Draper:2011aa}. This problem can be circumvented to some extent by either extending the gauge groups of the messenger sector \cite{Krauss:2013jva,Kahn:2013pfa} or by adding superpotential interactions between the matter and messenger fields \cite{Shadmi:2011hs,Evans:2011bea,Jelinski:2011xe,Evans:2012hg,Albaid:2012qk,Abdullah:2012tq,Perez:2012mj,Endo:2012rd}.

We are going the second way. In this work we propose a small extension of the minimal GMSB where one Higgs doublet interacts in the superpotential with two messenger fields. In addition, conformal sequestering with negative anomalous dimension is used to suppress the gaugino masses and A-terms.  We find that this model has a generic focus point. The simplicity of our model is a main improvement compared to previous attempts to combine GMSB and FP SUSY \cite{Agashe:1999ct}. So far these models have been very baroque and needed a complicated SUSY breaking mechanism. We will see that in the model presented here the fine-tuning issue is significantly alleviated compared to the minimal GMSB and the model remains natural. Using a precise, numerical setup we find that this model has a fine tuning of about 1000.

%

This paper is organized as follows. In section \ref{sec:model}, we present the details of our model and  derive the analytic solution for FP SUSY.  In section \ref{sec:numerical}, we consider the numerical studies of this model. The corresponding fine-tuning measure and phenomenology is discussed in details. We conclude in section~\ref{sec:conclusion}. The appendix contains two parts. In \ref{sec:convention}, the conventions and one-loop RGEs are given. In \ref{sec:derivation}, we derive the important formula which plays a crucial role in determining FP SUSY. 

\section{Focus Points SUSY in Yukawa Mediation}
\label{sec:model}
\subsection{Model Description}
In this paper, we propose an economic and complete model to achieve FP SUSY in GMSB. The messenger sector of our model
consists of a pair (${\bf 5}$, ${\bf \bar{5}}$) and an singlet under $SU(5)$. Thus, the gauge coupling unification
is preserved, and there is no Landau pole below the unification scale
because of the small messenger sector.
The messenger fields and their quantum numbers with
respect to $SU(3)_C \times SU(2)_L \times U(1)_Y$ are summarized in Tab.~\ref{table:messenger}.
\begin{table}[!htbp]
\label{table:messenger}
\centering
 \begin{tabular}{|l|c|c|c|}
  \hline
  & $SU(3)_C$ & $SU(2)_L$ & $U(1)_Y$ \\ \hline
  $ \Phi_1$ & $1$ &$2$ & $\frac{1}{2}$\\
  ${\tilde{\Phi}}_1$ & $1$ & $2$ & $-\frac{1}{2}$\\
  $ \Phi_2$ & $ 3 $ & $ 1 $ & $-\frac{1}{3}$\\
  ${\tilde{\Phi}}_2$ & $ \bar3 $ & $ 1 $ & $\frac{1}{3}$\\
  ${\tilde{\Phi}}_3$ & $ 1 $ & $ 1 $ & $0$\\ \hline
 \end{tabular}
   \caption{Representations of the messenger fields under the Standard Model gauge group.}
\end{table}
In the matter sector we have the common superfields of the MSSM. Their superpotential is
\begin{equation} \label{eq:W}
W_{MSSM} = Y_u U Q H_u + Y_d D Q H_d + Y_e E L H_d +  \mu H_u H_d \ .
\end{equation}
In addition, we introduce an interaction between $H_u$ and two messenger fields. The superpotential terms involving the messengers are
\begin{align}
W_{H\Phi} = X\,\Phi_i\tilde\Phi_i + \lambda_u H_u\Phi_3\tilde\Phi_1 \, .
\label{eqn:model}
\end{align}
Finally, SUSY is broken by some strong interactions in the hidden sector which we leave unspecified. These interactions cause a VEV for $X$ in its scalar and auxiliary components
\begin{equation}
X \to M_m + \theta^2 F \,,
\end{equation}
and we define  $\Lambda \equiv F/M_m$. The soft gaugino masses are created by one-loop interactions with the messenger and expected to be
$O\left(g_i^2/16\pi^2 \Lambda\right)$. In absence of any superpotential terms between messenger and MSSM fields, the
squared SUSY breaking soft scalar mass terms arise at the two-loop level and are generically of $O\left((g_i^2/16\pi^2)^2 \Lambda^2\right)$  \cite{Giudice:1998bp}. In addition to the common contributions from the gauge interactions with the messengers the soft-term for $H_u$ receives contributions  proportional to some power of $\lambda_u$  from the interaction given in Eq.~(\ref{eqn:model}). These contributions appear at one- and two-loops. In order to suppress the negative
one-loop corrections a large mediation scale of $10^8$~GeV is needed \cite{Craig:2012xp}. At two-loop $m_{H_u}^2$ receives a shift of the form
\begin{align}
\Delta m_{H_u}^2\sim \lambda_u^4-\lambda_u^2g^2~.~\,
\label{eqn:extra}
\end{align}
A precise expression for $\Delta m_{H_u}^2$ will be derived in Sec.~\ref{sec:Focus}. The FP SUSY requires $m_{H_u}^2$ to be comparable with squarks soft terms, i.e., a sizable positive $\Delta m_{H_u}^2$ is needed. In the messenger sectors larger
than the one discussed here, the superpotential interactions between the Higgs fields and messengers charged under $SU(3)_C$ might be allowed. However, those terms would cause negative contribution $\sim \lambda^2 g_3^2$. This make the minimal model even more attractive.

Here is a comment on the $A$-terms at place. The extra interaction between Higgs and messenger superfields is often used to generate  large A-terms as well. This enhances the Higgs mass and improves the fine-tuning. This setup has been already widely studied in the literatures, see e.g. Refs.~\cite{Kang:2012ra, Abdullah:2012tq, Byakti:2013ti, Craig:2012xp, Craig:2013wga, Evans:2013kxa, Calibbi:2013mka, Jelinski:2013kta, Galon:2013jba, Fischler:2013tva, Knapen:2013zla, Brummer:2013upa}. However, in this paper we assume that the  gaugino masses and all $A$-terms are suppressed through conformal sequestering as discussed below. Small gaugino masses are necessary to obtain a SUSY focus point and to reduce the fine-tuning as well. To obtain a focus point the gaugino contributions should be suppressed compared to the sfermion contributions. However, this can't be achieved in the minimal gauge mediation where the gaugino masses are of the same order as sfermion masses. Thus, we
consider the conformal sequestering in which gaugino masses are suppressed compared to sfermion masses. In conformal sequestering the gaugino masses are relatively light compared to the other masses because of large negative anomalous dimensions. We explain this in detail in Sec.~\ref{sec:conformal}.

In principle one could keep the $A$-terms large using conformal sequestering while only suppressing the gaugino masses. If $A_t$ would not be suppressed, the model will become FP SUSY with large A-term, which greatly improves the fine-tuning since it is easier to obtain $m_h\simeq 125$~GeV. We checked and found FP SUSY also including $A_t$. However, this choice is not natural because gaugino masses and $A_t$ should be treated at the same status. Therefore, we have not investigated this possibility further. As a result, the maximal mixing scenario could not be achieved and stop will be very heavy in order to satisfy $m_{h}=125$GeV.  In our setup stop masses of several TeV are needed and the overall fine-tuning is around $3000$, which is well accepted \cite{Feng:2013pwa,Craig:2013cxa} and a big improvement compared to the minimal GMSB.

\subsection{Analytical Derivation of Focus Point SUSY}
\label{sec:Focus}
The soft spectra of the model under considerations can be easily computed via the general formula given in Ref.~\cite{Evans:2013kxa}. Applied to our messenger sector and the interaction given in Eq.~(\ref{eqn:model}) the soft-breaking masses for all scalars are
\begin{align}
\tilde{m}^2_{H_{d}}&=\tilde{m}^2_L ~,~\, \nonumber\\
\tilde{m}^2_{H_{u}}&=n_{5}\frac{3}{10} \left({\mbox a}_1^2+5 {\mbox a}_2^2\right) \Lambda ^2
   f\left(\frac{\Lambda }{M_m}\right)+\left(-\frac{3}{5} {\mbox a}_1-3 {\mbox a}_2\right) \Lambda ^2 n_5 \alpha
   _{\lambda }+\Lambda ^2 \left(n_5^2+3 n_5\right) \alpha _{\lambda }^2  ~,~\, \nonumber\\
\tilde{m}^2_Q&=n_{5}\frac{1}{30} \left({\mbox a}_1^2+45 {\mbox a}_2^2+80 {\mbox a}_3^2\right) \Lambda ^2
   f\left(\frac{\Lambda}{M_m}\right)- n_{5} Y_t\alpha _{\lambda }\Lambda ^2 ~,~\, \nonumber\\
\tilde{m}^2_U&=n_{5}\frac{8}{15} \left({\mbox a}_1^2+5 {\mbox a}_3^2\right) \Lambda ^2
   f\left(\frac{\Lambda }{M_m}\right)- 2 n_{5}Y_t\alpha _{\lambda }\Lambda ^2 ~,~\, \nonumber\\
\tilde{m}^2_D&= n_{5}\frac{2}{15} \left({\mbox a}_1^2+20 {\mbox a}_3^2\right) \Lambda ^2
   f\left(\frac{\Lambda }{M_m}\right) ~,~\, \nonumber\\
\tilde{m}^2_L&=n_{5}\frac{3}{10} \left({\mbox a}_1^2+5 {\mbox a}_2^2\right) \Lambda ^2
   f\left(\frac{\Lambda }{M_m}\right)  ~,~\, \nonumber\\
\tilde{m}^2_E&=n_{5}\frac{6}{5} {\mbox a}_1^2 \Lambda ^2 f\left(\frac{\Lambda }{M_m}\right) ~,~\, \nonumber\\
A_t&=-\alpha _{\lambda }\Lambda  ~,~\, \nonumber\\
M_i&=g\left(\frac{\Lambda}{M_m}\right) {\mbox a}_i\Lambda ~.~\,
\label{eqn:softspectra}
\end{align}
Here, we used  ${\mbox a}_i = g_i^2/16\pi^2$ ($i=1,2,3$), $\alpha_\lambda = \lambda_u^2/16\pi^2$ and $n_5$ is the messenger index of the 5-plets.  $f$ and $g$ are loop-functions which can be found in Ref.~\cite{Giudice:1998bp}. $g\sim f \sim 1$ holds in the limit $M_m \gg \Lambda$.

As we have mentioned before, we are going to suppress A-term usually generated by Yukawa mediation and also gaugino masses by conformal sequestering. Thus, these contributions can be ignored in our analytical attempts to solve the RGEs. In addition, we neglect all Yukawa couplings except the top quark Yukawa coupling.  So the simplified limit is
\begin{align}
A_i,\ M_i,\ Y_{b},\ Y_{\tau}\rightarrow 0 \, .
\label{eqn:simplify}
\end{align}
To determine the focus point, $\tilde{m}_{H_{u}}^2$ at the weak scale should be written as a function of soft spectra at the conformal scale. Actually, this could be easily obtained when we use the one-loop RGEs given in appendix~\ref{sec:convention}. In the limit (\ref{eqn:simplify}) the RGEs for the Higgs and stop soft-terms are
\begin{align}
\frac{d\tilde{m}^2_{H_{u}}}{dt} & =-3 Y_t \left(\tilde{m}_{H_u}^2+\tilde{m}_{Q}^2+\tilde{m}_{U}^2\right) ~,~\,
\label{eqn:rge1} \\
\frac{d\tilde{m}^2_Q}{dt} & =-Y_t \left(\tilde{m}_{H_u}^2+\tilde{m}_{Q}^2+\tilde{m}_{U}^2\right) ~,~\,  \\
\frac{d\tilde{m}^2_U}{dt} & =-2Y_t \left(\tilde{m}_{H_u}^2+\tilde{m}_{Q}^2+\tilde{m}_{U}^2\right) ~.~\,
\label{eqn:rge2}
\end{align}
The $\beta$-functions of all other soft-scalar masses vanish in the limit (\ref{eqn:simplify}).
Eqs.~(\ref{eqn:rge1})--(\ref{eqn:rge2}) can be solved simultaneously and we find
\begin{equation}
\tilde{m}^2_{H_{u}}[t] =\frac{1}{2} (\tilde{m}_{H_{u}}^2[0](\mathcal{I}+1)+(\mathcal{I}-1)
   (\tilde{m}_{Q}^2[0]+\tilde{m}_{U}^2[0]))  ~,~\,
\end{equation}
with
\begin{equation}
\mathcal{I} =\exp\left(-6 \int_0^t Y_t[t']dt'\right) ~,~\,
\end{equation}
where $\mathcal{I}$ is computed in appendix~\ref{sec:derivation}. The FP SUSY is found at $\tilde{m}^2_{H_{u}}[t]=0$, which requires
\begin{equation}
\frac{\tilde{m}^2_{H_{u}}[0]}{\tilde{m}_{Q}^2[0]+\tilde{m}_{U}^2[0]}=\frac{1-\mathcal{I}}{1+\mathcal{I}}.
\label{eq:cond}
\end{equation}
\begin{figure}[!htbp]
\begin{center}
\includegraphics[scale=.71]{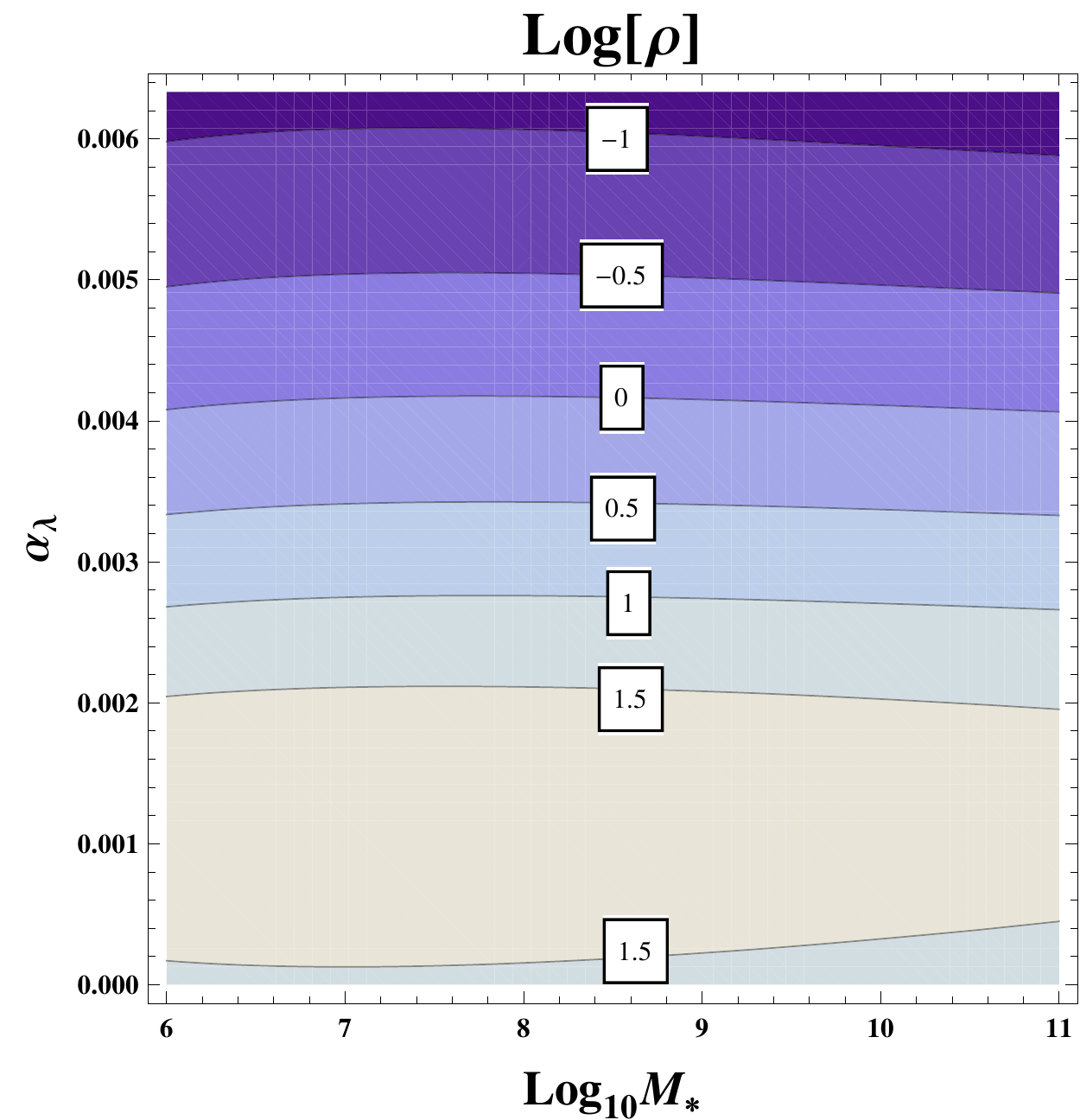}
\caption{In this figure, we take $\alpha_{em}^{-1}=127.931993$,
$\alpha_{s}=0.1720$, $m_{Z}=91.1876$, top quark pole mass $m_{t}=172.9$, and $\tan\beta=10$. Meanwhile, the high scale input includes $n_{5}=1$. In addition, we point out that
for arbitrary $n_{5}<5$, the focus point SUSY is generic.
}
\label{fig:FocusPoint}
\end{center}
\end{figure}
It has been proven in Ref.~\cite{Agashe:1999ct} that the minimal gauge mediation cannot provide the required ratio. The reason is that $\tilde{m}^2_{H_{u}}$ is significantly smaller than squarks soft-term in the minimal gauge mediation because of the dominant contributions from the strongly interacting messengers. However, through the Yukawa mediation, the extra two-loop positive contribution for $\tilde{m}^2_{H_{u}}$ and negative contributions to $\tilde{m}^2_{Q}$/$\tilde{m}^2_{U}$ are combined to yield a realistic model with focusing behavior. For convenience, we define the required and actually value of eq.~(\ref{eq:cond}) as
\begin{eqnarray}
y_{req}=\frac{1-\mathcal{I}}{1+\mathcal{I}}\,,\hspace{0.5cm}
y_{act}=\frac{\tilde{m}^2_{H_{u}}[0]}{\tilde{m}_{Q}^2[0]+\tilde{m}_{U}^2[0]}\,,\hspace{0.5cm}
\end{eqnarray}
while the ratio is given by
\begin{equation}
\rho=\frac{y_{req}}{y_{act}} \,.
\end{equation}
It is easy to see from Fig.~\ref{fig:FocusPoint} that $\log\rho\simeq0$ can be naturally satisfied for moderate values of $\alpha_\lambda$ and a wide range in $M_*$.

\subsection{Reducing Gaugino Mass Fine-tuning via Conformal Sequestering}
\label{sec:conformal}
In the previous section \ref{sec:model} the suppressed gaugino masses and A-terms have been assumed in order to generate FP SUSY. Here, we present a possible origin of  this suppression. To this end we follow previous studies of conformal sequestering in terms of the effective field theory below the messenger scale $M_m$ \cite{Cohen:2006qc,Roy:2007nz,Murayama:2007ge,Shirai:2008qt,Asano:2008qc,Perez:2008ng}. In this setup, the visible and hidden sectors are coupled through irrelevant operators in the K\"ahler potential. We summarize here the main idea and refer the interested reader for many more details to Ref.~\cite{Murayama:2007ge}.

In gauge mediation the gaugino and scalar masses are generated after integrating out the messenger multiplets at
 respectively one- and two-loop level. The effective interactions for the gauge and matter multiplets in the MSSM with a singlet in the hidden sector $X$ are
\begin{equation}
\label{eq:Leff}
\mathcal{L}_{\rm eff}=
  \left[\int d^2\theta \sum_{a=1}^3 \frac{1}{2} c^a_{\lambda} \frac{X}{M_m}{\mathcal{W}}^{a\alpha}{\mathcal{W}}^{a}_{\alpha} +h.c.\right]-
  \int d^4\theta \sum_{\tilde{f}} c^{\tilde{f}}_{m^2} \frac{X^\dagger X}{M_m^2} \tilde{f}^\dagger \tilde{f}.
\end{equation}
Here, ${\mathcal{W}}^{a\alpha}$ (with $a=1,2,3$) are the field strength superfields for the SM gauge sector and $\tilde f\in \{q,u,d,l,e,H_u,H_d\}$. The coefficients $c^a_\lambda$ appear at one loop and $c^{\tilde{f}}_{m^2}$ at two loop. The precise definitions of these coefficients are given in Ref.~\cite{Martin:1996zb}.

When the hidden sector enters the conformal regime at some scale $M_{*}$ the two terms in Eq.~(\ref{eq:Leff}) receive large corrections from wave function renormalization. The effective Lagrangian is then given at some renormalization scale $\mu_R$ (with $\mu_R < M_m$) by
\begin{align}
{\cal L}_{\rm eff}=
\left[\int d^2\theta \sum_{a=1}^{3} \frac{1}{2}c_{g} Z_X^{-1/2} \frac{X}{M_m}{\cal W}^{a\alpha}{\cal W}^{a}_{\alpha} +h.c.\right]-
\int d^4\theta \sum_{\tilde{f}} c^{\tilde{f}}_{m^2}
Z_X^{-1}  Z_{|X|^2}
 \frac{X^\dagger X}{M_m^2} \tilde{f}^\dagger \tilde{f} ~.~\,
 \label{eqn:effective}
\end{align}
From this equation it can be seen that the wave function renormalization constants $Z_X^{-1/2}$ and $Z_{|X|^2}$ can be used to suppress either the scalar or the gaugino soft masses. At the conformal scale $M_*$ the renormalization constants are given by
 \begin{equation}
 Z_X[0]=\left(\frac{M_m}{M_*}\right)^{3 R(X)-2}\,~,~
 Z_{|X|^2}[0]=\left(\frac{M_*}{M_m}\right)^{\gamma}\,.
 \end{equation}
Here, $[0]$ indicates the quantities which are evaluated at the conformal scale. In the Higgs sector, the $\mu$ term, $B_{\mu}$ and $A_{H_d}$ cannot be generated at the messenger scale because the messenger couple only to $H_u$. Thus, the effective Lagrangian at conformal scale is given by
\begin{align}
{\cal L}_{\rm eff}= -\left[\int d^4\theta c_{A_u} Z_X^{-1/2} \frac{X}{M_m}{H_u H_u^{\dagger}}\right]-
\int d^4\theta Z_X^{-1}  Z_{|X|^2}  \frac{X^\dagger X}{M_m^2} \left(H_u H_u^{\dagger}+H_d H_d^{\dagger}\right)~.
 \label{eqn:effective2}
\end{align}
 Using Eqs.~(\ref{eqn:effective}) to (\ref{eqn:effective2}), the soft SUSY spectra at the conformal scale $M_*$ can be related to those at messenger scale $M_m$ via
 \begin{equation}
 M_i[0]=Z_X^{-1/2}M_i[t_m]\,,\hspace{0.5cm}
 A_t[0]=Z_X^{-1/2} A_t[t_m]\,,\hspace{0.5cm}
 m_f^2[0]=Z_X^{-1}Z_{|X|^2} m_{\tilde f}^2[t_m] \, ,
 \end{equation}
with $t_m = 2\log(\frac{M_*}{M_m})$.
For $\gamma>0$ and $R(X)>3/2$  the conventional conformal sequestering is achieved and one finds $m_{\tilde f}^2\approx 0$. However, it has been shown in Ref.~\cite{Poland:2011ey} that this type of conformal sequestering is severely constrained by stringent bounds on $\gamma$ from internal consistency of the hidden sector superconformal field theory (SCFT). For $\text{dim}(X)\simeq1$ self-consistent condition requires $\gamma=\text{dim}(X X^{\dagger})-2\text{dim}(X)<0$. This forbids the
positive anomalous dimensions and as consequence the sfermion and Higgs masses are enhanced compared to the gaugino masses.

We make use of this suppression of the gaugino masses and A-terms relative to the sfermion and Higgs masses in the case of $\gamma<0$. For this purpose we parametrize
 \begin{equation}
 Z_X^{-1/2}=\eta\nonumber\,,\hspace{0.5cm}
 Z_{|X|^2}=\frac{1}{\epsilon^2}\nonumber\,,\hspace{0.5cm}
 \Lambda_a=\frac{\eta}{\epsilon} \Lambda \, .
 \label{eqn:definition}
 \end{equation}
 Using these definitions we can relate $\Lambda$ and $M_m$ to $\Lambda_a$ and $M_*$ via
\begin{align}
\frac{\Lambda}{M_m} =\left(\frac{\epsilon\Lambda_a}{\eta M_*}\right)^2 ~.~
\end{align}
We finally end up with the following boundary conditions for the soft masses in our GMSB version of the MSSM
\begin{align}
\tilde{m}^2_{H_{d}}[0]&=\tilde{m}^2_L[0] ~,~ \nonumber\\
\tilde{m}^2_{H_{u}}[0]&=n_{5}\frac{3}{10} \left({\mbox a}_1[0]^2+5 {\mbox a}_2[0]^2\right) \Lambda_a ^2
   f\left(\frac{\eta^2\Lambda^2 }{\epsilon^2M_*^2}\right)+\left(-\frac{3}{5} {\mbox a}_1[0]-3 {\mbox a}_2[0]\right) \Lambda_a ^2 n_5 \alpha
   _{\lambda }+\Lambda_a ^2 \left(n_5^2+3 n_5\right) \alpha _{\lambda }^2 ~,~ \nonumber\\
\tilde{m}^2_Q[0]&=n_{5}\frac{1}{30} \left({\mbox a}_1[0]^2+45 {\mbox a}_2[0]^2+80 {\mbox a}_3[0]^2\right) \Lambda_a ^2
   f\left(\frac{\eta^2\Lambda_a^2 }{\epsilon^2M_*^2}\right)- n_{5} Y_t[0]\alpha _{\lambda }\Lambda_a ^2 ~,~ \nonumber\\
\tilde{m}^2_U[0]&=n_{5}\frac{8}{15} \left({\mbox a}_1[0]^2+5 {\mbox a}_3[0]^2\right) \Lambda_a ^2
   f\left(\frac{\eta^2\Lambda_a^2 }{\epsilon^2M_*^2}\right)- 2 n_{5}Y_t[0]\alpha _{\lambda }\Lambda_a ^2 ~,~ \nonumber\\
\tilde{m}^2_D[0]&= n_{5}\frac{2}{15} \left({\mbox a}_1[0]^2+20 {\mbox a}_3[0]{}^2\right) \Lambda_a ^2
   f\left(\frac{\eta^2\Lambda_a^2 }{\epsilon^2M_*^2}\right) ~,~ \nonumber\\
\tilde{m}^2_L[0]&=n_{5}\frac{3}{10} \left({\mbox a}_1[0]^2+5 {\mbox a}_2[0]^2\right) \Lambda_a ^2
   f\left(\frac{\eta^2\Lambda_a^2 }{\epsilon^2M_*^2}\right)  ~,~ \nonumber\\
\tilde{m}^2_E[0]&=n_{5}\frac{6}{5} {\mbox a}_1[0]2 \Lambda_a ^2 f\left(\frac{\eta^2\Lambda_a^2 }{\epsilon^2M_*^2}\right) ~,~ \nonumber\\
A_t[0]&=\epsilon\alpha _{\lambda }[0]\Lambda_a  ~,~ \nonumber\\
M_i[0]&=\epsilon{\mbox a}_i[0]\Lambda_a g\left(\frac{\eta^2\Lambda_a^2 }{\epsilon^2M_*^2}\right)~.~
\label{eqn:softtransformation}
\end{align}
To sum up, the free parameters of this model are $\{\eta,\;\epsilon,\;n_5,\;\Lambda_a,\;\lambda_u,\;M_*,\;\tan\beta,\;\text{sign}(\mu)\}$. Since $\eta$ only enters in the loop function $f(x)$ and $g(x)$ one can impose $\eta=\epsilon$ for simplicity to remove one degree of freedom. In total, there are six free parameters and one sign in this model
\begin{align}
 \{\epsilon,\;n_5,\;\Lambda_a,\;\lambda_u,\;M_*,\;\tan\beta,\;\text{sign}(\mu)\} \, .
\end{align}
Without leading to confusion, the parameter $\Lambda_a$ could is identified with $\Lambda$ after imposing $\eta=\epsilon$, and  
\begin{align}
M_m\Lambda=M_*^2
\end{align}
\section{Naturalness and Phenomenology}
\label{sec:numerical}
We present in this section our numerical results for the fine tuning in this model and comment on some phenomenological features. For this purpose, we have implemented this model in the Mathematica package {\tt SARAH}~\cite{Staub:2008uz,Staub:2009bi,Staub:2010jh,Staub:2012pb,Staub:2013tta}. {\tt SARAH} has been used to create a {\tt SPheno}\cite{Porod:2003um,Porod:2011nf}  version for the MSSM with the new boundary conditions for the soft-terms at the messenger scale. The {\tt SARAH} generated {\tt SPheno} version calculates the mass spectrum with the same precision as {\tt SPheno 3.2.4} but includes also routines to calculate the fine tuning according to Eq.~(\ref{eq:FT}). In our case the free parameters which influence the fine tuning are slightly different compared to Ref.~\cite{Evans:2013kxa} since we have suppressed all one-loop contributions to the soft-masses at the messenger scale to have a sufficiently large $m_{H_u}^2$. Thus, we calculate the fine tuning with respect to
\begin{eqnarray}
&a\in \{ \Lambda,\; \lambda_{u},\; \mu,\;g_{3},\;y_t\} \, .&
\label{eq:ftpara}
\end{eqnarray}
The null results from SUSY searches at the LHC put severe limits on the allowed masses of the gluino and of the squarks of the first two generations \cite{Aad:2012fqa,Chatrchyan:2012jx}. These limits can roughly be summarized to: (i) $m_{\tilde g}\gtrsim 1.5$ TeV (for $m_{\tilde g}\simeq m_{\tilde q}$), (ii) $m_{\tilde g}\gtrsim1$ TeV (for $m_{\tilde g}\ll m_{\tilde q}$). In our case the gluino is always much lighter than the squarks because of the suppression from conformal sequestering. Thus, we impose the constraint $m_{\tilde g}\gtrsim1$~TeV in addition to $123<m_{h}<129$~GeV.

To check the overall fine tuning in our model we have performed a random scan of $2\times10^5$ points
in the following parameter ranges
\begin{eqnarray}
10^4 {\rm GeV}& <\Lambda_{a}< & 10^6 {\rm GeV} ~,~ \nonumber \\
10^6 {\rm GeV}& <M_{*}<& 10^{11} {\rm GeV} ~,~ \nonumber\\
0&<\lambda_{u}<&1 ~,~ \nonumber\\
0.1&<\ \epsilon\ <&1 ~.~
\label{eq:ranges}
\end{eqnarray}
The other parameters have been fixed to $\tan \beta=10 $, $n_{5}=1$, $n_{10}=0$ and $\text{sign}(\mu)=1$. The parameter scans have been performed with {\tt SSP} \cite{Staub:2011dp}.  The point with the best fine tuning satisfying the constraints on the SUSY and Higgs masses has a fine tuning of $\Delta = 3117$, see for more details Table~\ref{table:modelsfull}. Thus, even if we keep
the mixing in the stop very small and need very heavy stops to obtain the correct Higgs, a fine tuning of about $3000$ is still very good compared the CMSSM expectations with $A_0=0$. Because the electroweak gauge symmetry will not be broken
at the weak scale if supersymmetry is not broken, the relevent parameters for fine-tuning measures
are $\mu$ and $\Lambda_{a}$. Also, the Yukawa couplings can be calculated in the concrete model building,
for example, string models.
 Thus, we can just consider $\mu$ and $\Lambda_{a}$ for fine-tuning study, and the corresponding
 fine tuning is reduced to $\Delta \sim 1000$. A benchmark point is given in Table~\ref{table:modelsmall}.

\begin{table}[t]
\begin{center}
\begin{tabular}{|c|c|c|c|c|c|c|}
\hline
$\Lambda_{a}$ (GeV), $M_{*}$ (GeV), $\lambda_u$, $\epsilon$ & $\Delta_{FT}$ & ${A_t}/M_S$ & $m_{\tilde g}$ (GeV) & $m_{\tilde t_{1}}$ (GeV) & $m_{\tilde \tau_{1}}$ (GeV) & $m_{\tilde \chi^{0}_{1}}$ (GeV)  \\
\hline
$8\times10^5$, $8.5\times10^8$, $0.57$, $0.39$ &$3117$&$-0.28$&$2333$&$4299$&$1573$&$436$  \\
\hline
\end{tabular}
\caption{The point with minimal fine tuning in our random scans. The parameter ranges have been chosen according to Eq.~(\ref{eq:ranges}).}
\label{table:modelsfull}
\end{center}
\end{table}

\begin{table}[t]
\begin{center}
\begin{tabular}{|c|c|c|c|c|c|c|}
\hline
$\Lambda_{a}$ (GeV), $M_{*}$ (GeV), $\lambda_u$, $\epsilon$ & $\Delta_{FT}$ & ${A_t}/M_S$ & $m_{\tilde g}$ (GeV) & $m_{\tilde t_{1}}$ (GeV) & $m_{\tilde \tau_{1}}$ (GeV) & $m_{\tilde \chi^{0}_{1}}$ (GeV)  \\
\hline
$8.7\times10^5$, $9.5\times10^{9}$, $0.71$, $0.27$ &$1127$&$-0.28$&$1831$&$3496$&$1692$&$327$  \\
\hline
\end{tabular}
\caption{Similar to table \ref{table:modelsfull}, but here we only consider the $\mu$ and $\Lambda_{a}$ for tuning measure.}
\label{table:modelsmall}
\end{center}
\end{table}

\begin{figure} [t]
\begin{center}
\includegraphics[width=0.48\textwidth]{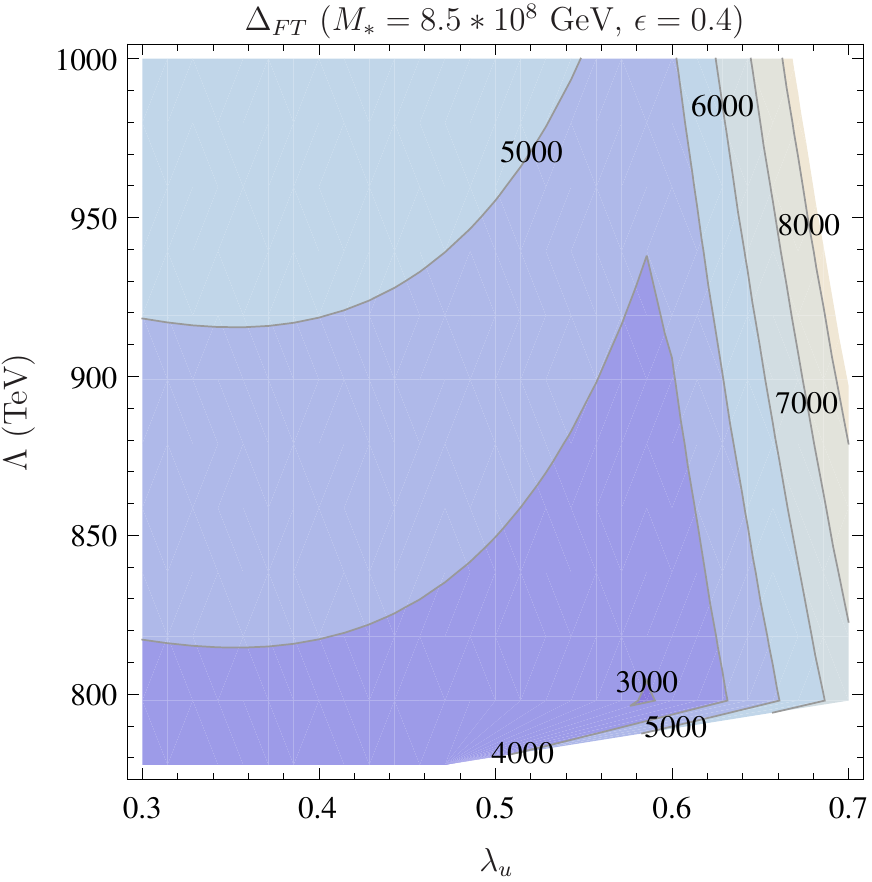}
\includegraphics[width=0.48\textwidth]{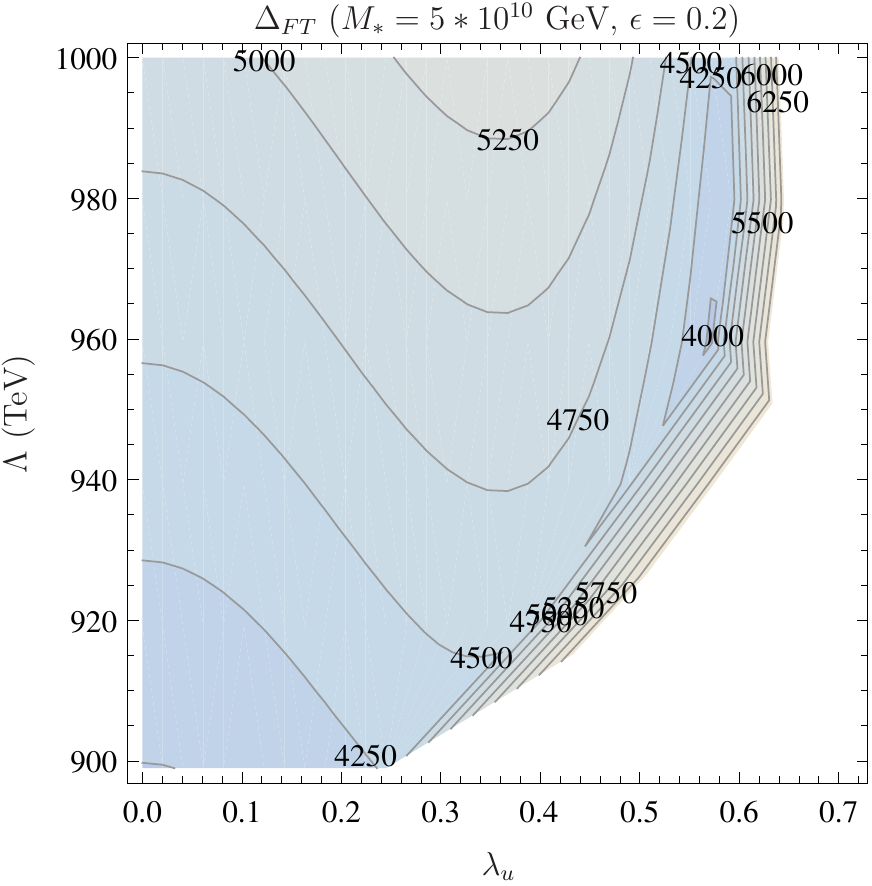}
\end{center}
\caption{Fine tuning in the  $(\lambda_{u},\Lambda_{a})$ plane for $\epsilon =0.4$, $M_{*}=8.5\times10^8$~GeV (left-panel) and  $\epsilon =0.2$, $M_{*}=5\times10^{10}$~GeV. The other parameters are fixed to $\tan \beta=10$, $n_{5}=1$, $n_{10}=0$ and $\text{sign}(\mu)=1$.}
\label{fig:FT_1}
\end{figure}

In Fig.~\ref{fig:FT_1}, we show the fine tuning in the $(\lambda_{u},\Lambda_{a})$ plane for two different combinations of $\epsilon$ and $M_*$: ($\epsilon =0.4$, $M_{*}=8.5\times10^{8}$~GeV) and ($\epsilon =0.2$, $M_{*}=5\times10^{10}$~GeV).
The behavior of the fine tuning can be summarized as follows
\begin{enumerate}
\item In the $(\lambda_{u},\Lambda_{a})$ space: for a given value of $M_{*}$ and $\epsilon$, increasing $\Lambda_{a}$ and $\lambda_{u}$ increases the overall fine tuning. The reason is that large $\Lambda_{a}$ and $\lambda_{u}$ increase $\delta m_{H_u}^2$, see Eqs.~(\ref{eqn:deltaHu}) and (\ref{eqn:extra}).
\item Small values of $\Lambda_{a}$ and $\lambda_{u}$: $\Delta$ is usually dominated by $\mu$. Since in these regions the RGE effects are most important, the contribution to the fine tuning of $\lambda_{u}$, which only affects the boundary conditions, is negligible. The important parameters are $\Lambda_a$ which sets the range of the RGE running and, even more important, the absolute value of $\mu$.
\item moderate $\Lambda_{a}$ and $\lambda_{u}$: the contribution from $\mu$ and $\Lambda_{a}$ are almost comparable.
\item large $\lambda_{u}$: if $\lambda_u$ becomes large it is always the biggest contributor to fine tuning measure independent of the value of $\Lambda_{a}$. This seems to contradict the requirement of FP SUSY, for which special $\lambda_u$ gives rise to the focusing behavior thus reducing the fine tuning. However, small changes in $\lambda_u$ lead to sizable changes in the Higgs soft parameter at the threshold scale. The problem would improve if $\lambda_u$ has a fixed point for a given $M_{*}$. In that case the fine tuning induced by $\lambda_u$ would be completely negligible. We leave this topic for future investigation and accept here the fine tuning with respect to $\lambda_u$. Once we eliminate the couplings $y_t$, $g_3$ and $\lambda_u$ from the fundamental parameters, the corresponding fine-tuning will become $1000$, which falls into the regime of natural SUSY.
\end{enumerate}

\begin{figure} [t]
\begin{center}
\includegraphics[width=0.48\textwidth]{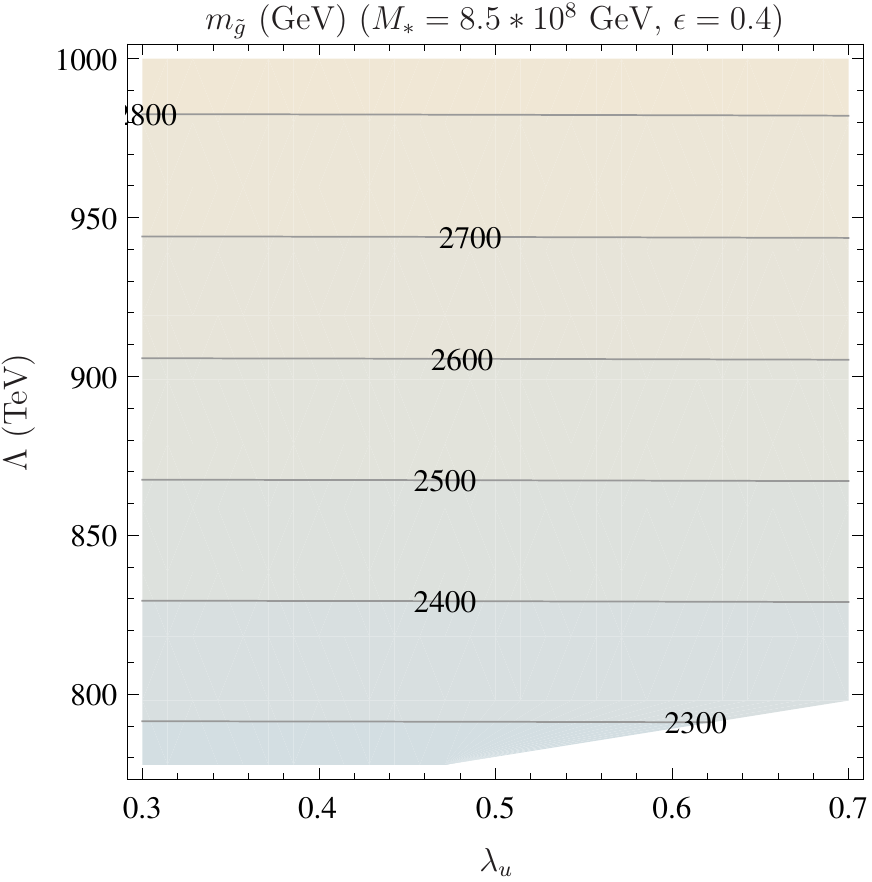}
\includegraphics[width=0.48\textwidth]{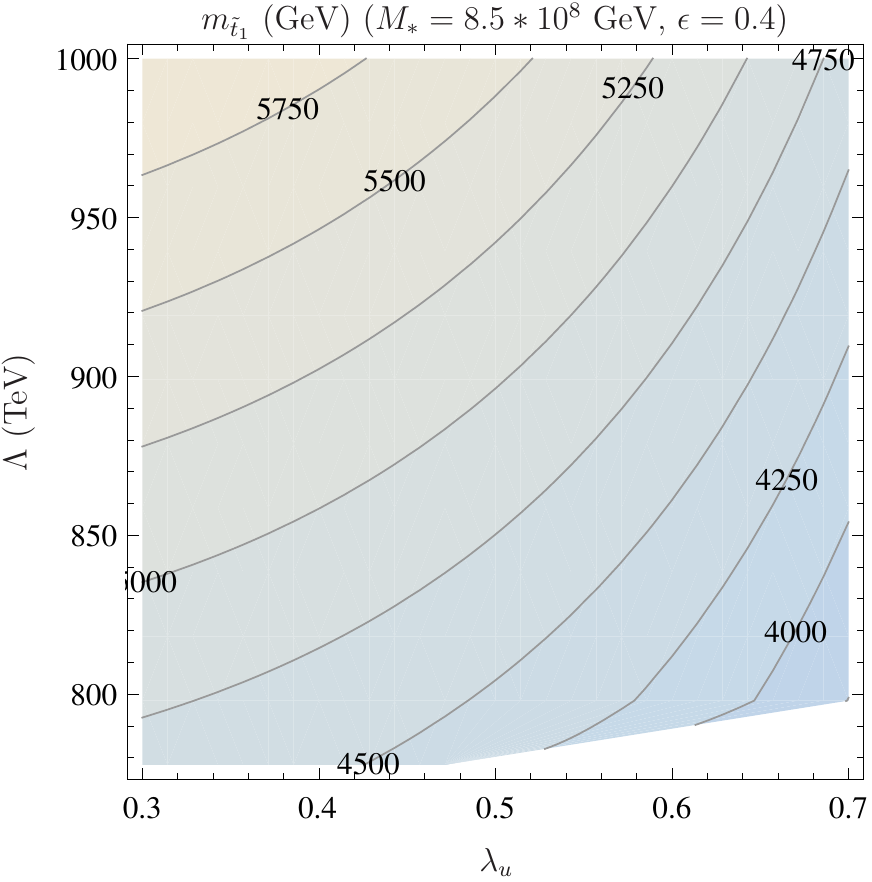}
\includegraphics[width=0.48\textwidth]{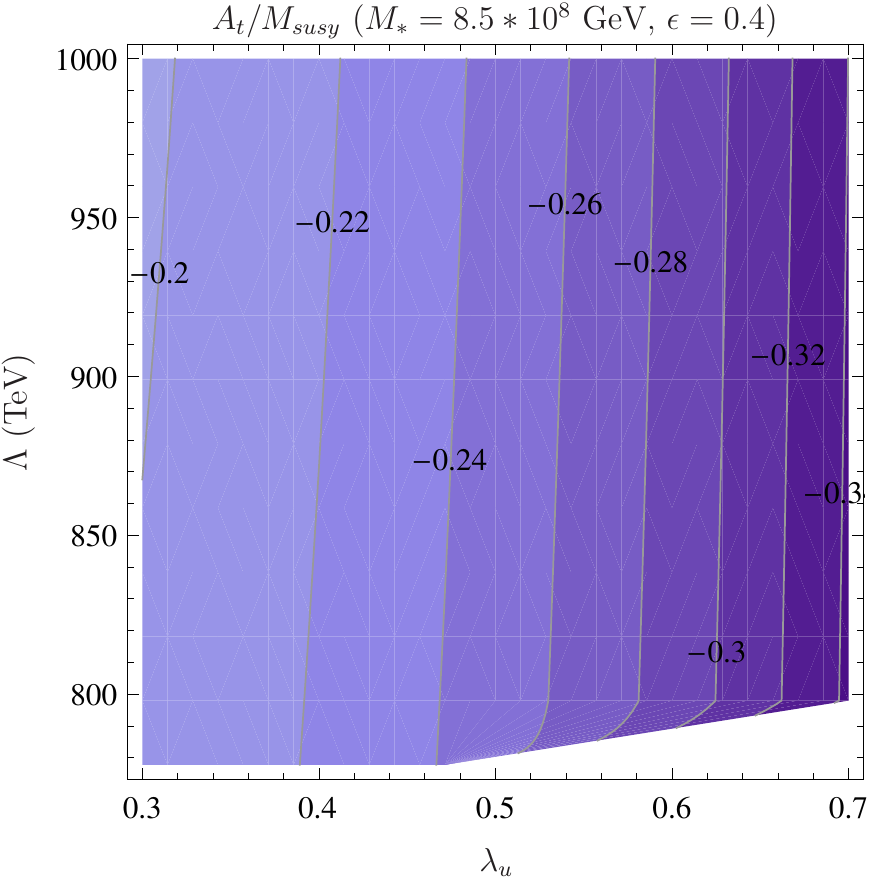}
\end{center}
\caption{Gluino mass $m_{\tilde g}$ (top-left), stop mass $m_{\tilde t_{1}}$ (top-right), and stop mixing ${A_t}/M_S$ (bootom) in the $(\lambda_{u},\Lambda_{a})$ plane for $M_{*}=8.5\times10^8$~GeV, and $\epsilon=0.4$. The other parameters are fixed to $\tan \beta=10$, $n_{5}=1$, $n_{10}=0$ and $\text{sign}(\mu)=1$.}
\label{fig:spectrum_1}
\end{figure}

\begin{figure} [t]
\begin{center}
\includegraphics[width=0.48\textwidth]{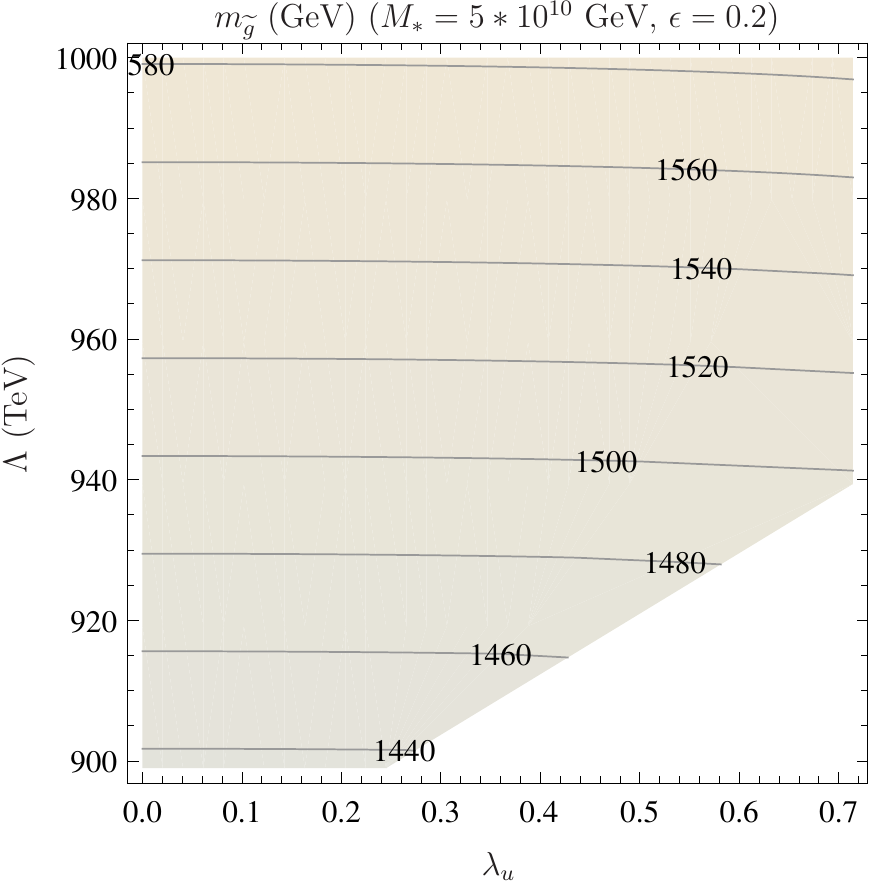}
\includegraphics[width=0.48\textwidth]{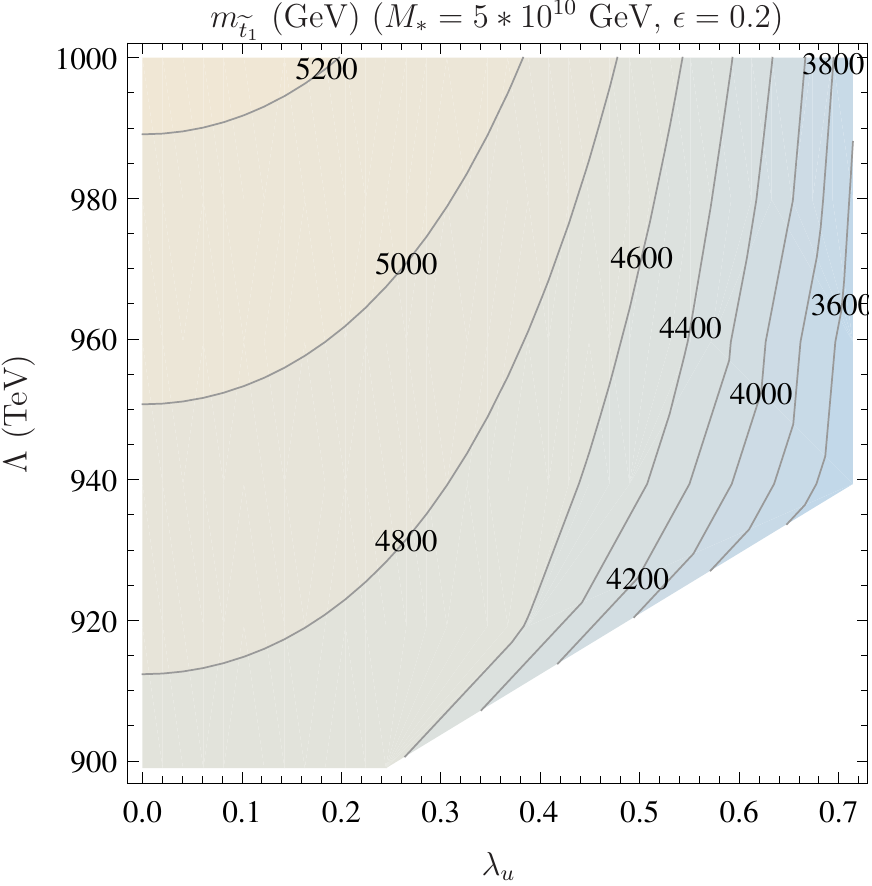}
\includegraphics[width=0.48\textwidth]{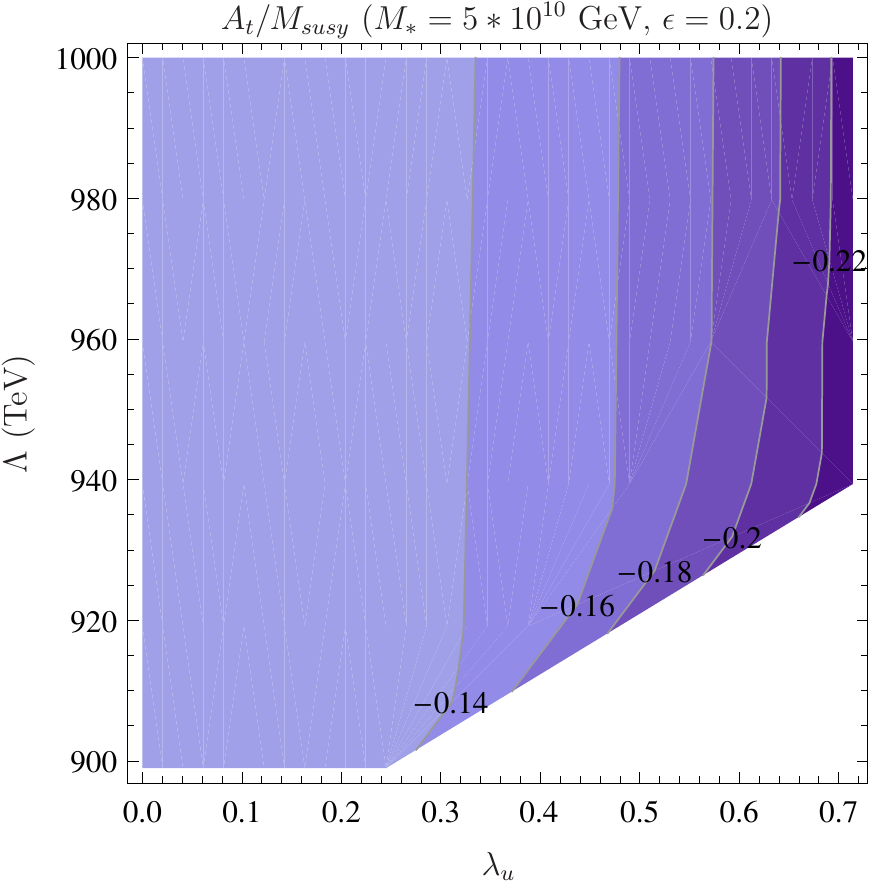}
\end{center}
\caption{Similar to figure ~\ref{fig:spectrum_1}, but for $M_{*}=5\times10^{10}$~GeV, $\epsilon=0.2$.}
\label{fig:spectrum_2}
\end{figure}

\begin{figure} [t]
\begin{center}
\includegraphics[width=0.48\textwidth]{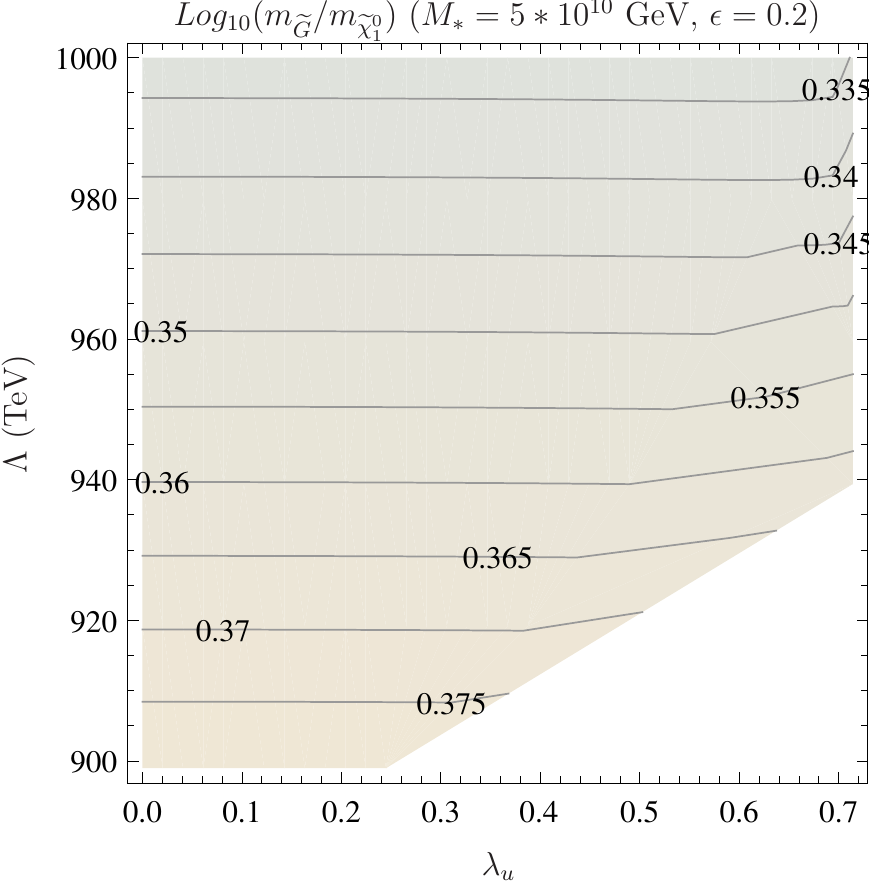}
\includegraphics[width=0.48\textwidth]{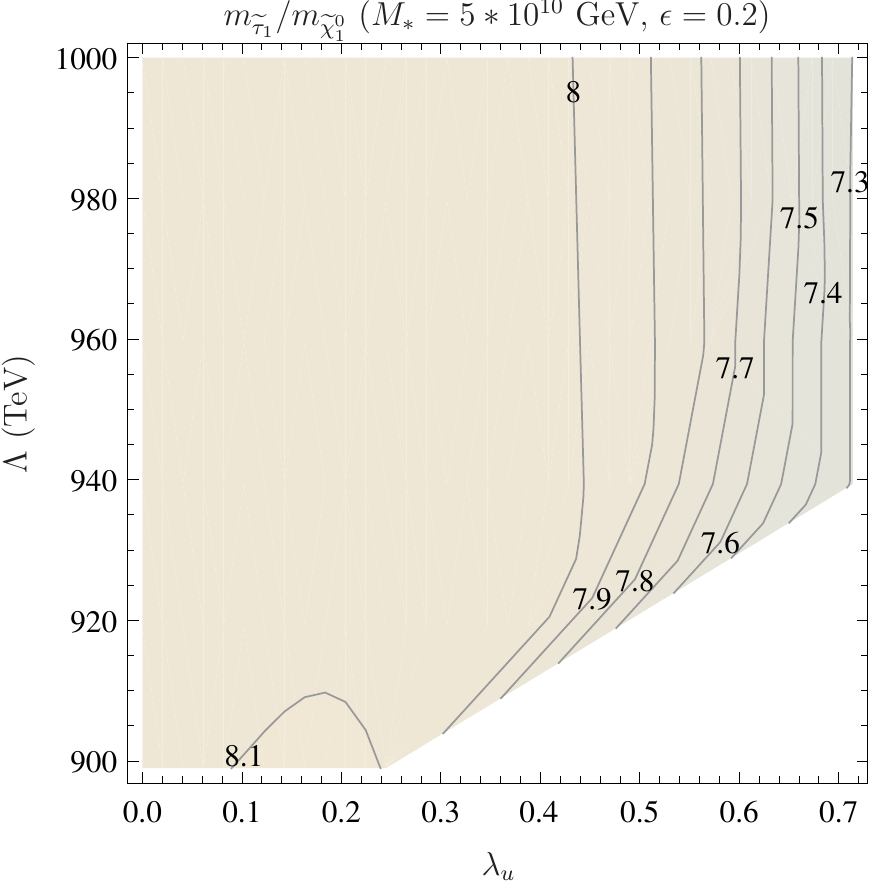}
\includegraphics[width=0.48\textwidth]{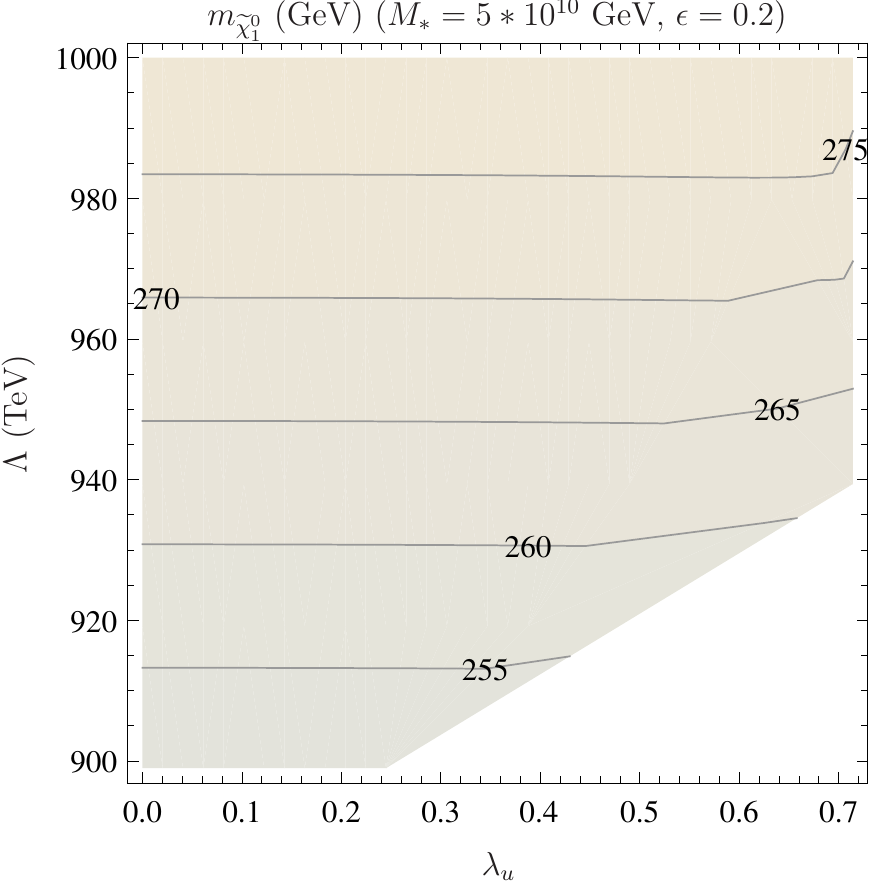}
\end{center}
\caption{Mass ratio of the light stau and the neutralino ($m_{\tilde \tau_{1}}/m_{\tilde \chi^{0}_{1}}$, on the top-left) , the mass ratio of the gravitino and neutralino ($Log_{10} m_{\tilde G}/m_{\tilde \chi^{0}_{1}}$, top-right) and absolute value of the neutralino mass ($m_{\tilde \chi^{0}_{1}}$, bottom) in the $(\lambda_{u},\Lambda_{a})$ plane. The other parameters are fixed to $M_*=5\times10^{10}$~GeV, $\epsilon=0.2$, $\tan \beta=10$, $n_{5}=1$, $n_{10}=0$, and $\text{sign}(\mu)=1$. }
\label{fig:spectrum_3}
\end{figure}

We shall finish some comments on the phenomenological aspects of the focus point in the presented model. For this purpose, we show in the Figs.~\ref{fig:spectrum_1}--\ref{fig:spectrum_3} the contours of relevant SUSY masses in the  $(\lambda_{u},\Lambda_{a})$ plane for the same combinations of $\epsilon$ and $M_*$ as in Fig.~\ref{fig:FT_1}. As it can be seen in Fig.~\ref{fig:spectrum_2}, the gluino for $\epsilon=0.2$ is well below 2~TeV and then within the reach of the next run of the LHC experiments. The stop is always in the multi TeV range and therefore out of reach. Hence, this model leads naturally to a split SUSY behavior which is widely discussed in literatures, see e.g. \cite{ArkaniHamed:2004yi,Giudice:2004tc,ArkaniHamed:2004yi,Bernal:2007uv,Harigaya:2013asa,Alves:2011ug}. In addition, we show the ratio ${A_t}/M_S$ which is always small in the entire range.

In Fig.~\ref{fig:spectrum_3}, we consider which parameters are interesting with respect to the dark matter properties of this model: the mass of the lightest neutralinos $m_{\tilde \chi^{0}_{1}}$, the mass ratio between stau and neutralino $m_{\tilde \tau_{1}}/m_{\tilde \chi^{0}_{1}}$ and the mass ratio of the Gravitino and neutralino $\text{log}_{10}\left(m_{\tilde G}/m_{\tilde \chi^{0}_{1}}\right)$. Here, we take again the same parameter space as in Fig.~\ref{fig:FT_1}. It is well known that the gravitino is usually the lightest supersymmetric particle (LSP) in conventional GMSB model with a mass of
\begin{equation}
m_{3/2}=\frac{F}{\sqrt{3}M_{\rm Pl}}~,~
\end{equation}
with $F=M_*^2$.
Here, $M_{\rm Pl}=2.4\times10^{18}$~GeV is the reduced Planck mass. One of the most intriguing feature is that at high $M_{*}$ scale, the LSP in mass spectrum will naturally become the neutralino which is the promising dark matter candidate which does not suffer from the cosmological gravitino problem \cite{Banks:1993en,Moroi:1993mb,Hall:1983iz,Baltz:2001rq,Kawasaki:2006gs,Kawasaki:2008qe,Staub:2009ww}. To demonstrate this feature, we show the mass ratio of the gravitino and neutralino. It can be seen clearly that the
neutralino is the LSP in the entire parameter region for $M_{*}\sim5\times10^{10}$ GeV. The mass of the neutralino is of the order of a few hundred GeV, i.e., of the typical range of WIMP (weakly interacting massive particle) candidate for dark matter.

We have only touched here the interesting phenomenological aspects of the model but concentrated on the fine tuning properties. A detailed discussion of the mass spectrum and the dark matter properties of a neutralino LSP will be given elsewhere.


\begin{figure} [t]
\begin{center}
\includegraphics[bb={157 465 478 695},scale=1.0]{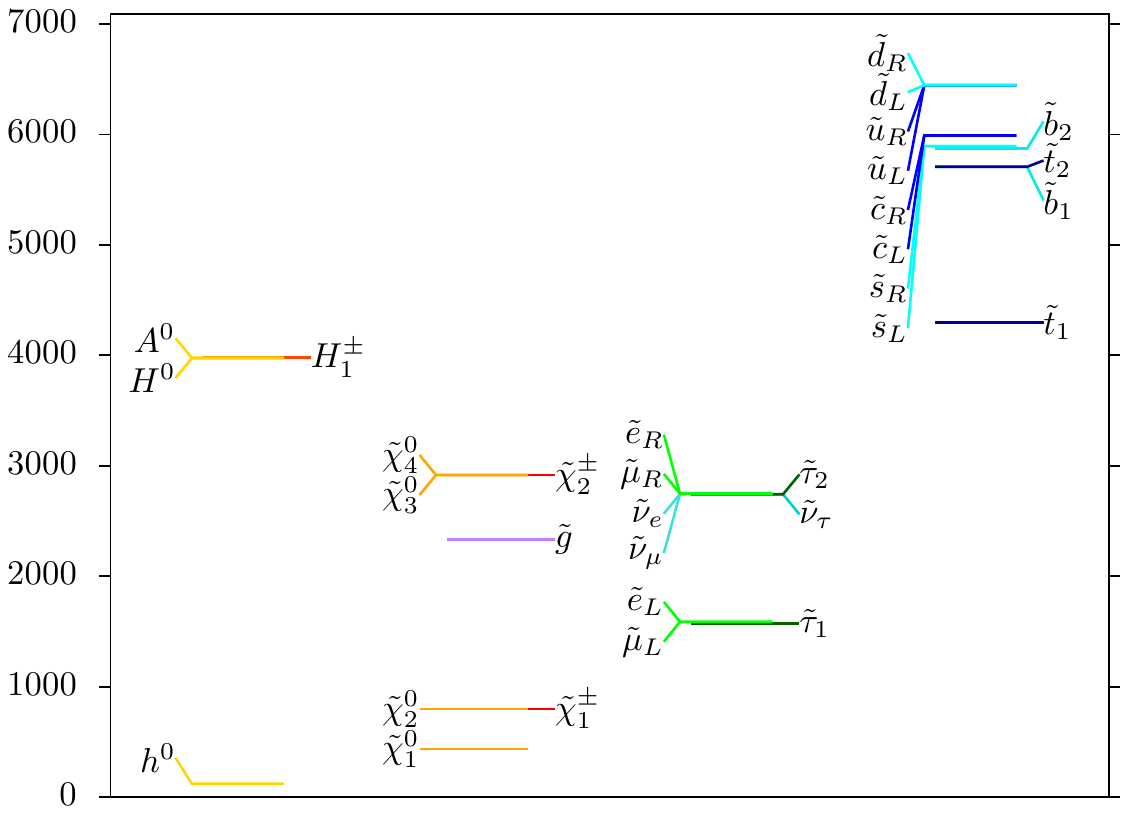}
\end{center}
\caption{Mass Spectrum in the Best Point, in this case the LSP is still gravitino as in conventional GMSB model.}
\label{fig:spectrum_best}
\end{figure}

\begin{figure} [t]
\begin{center}
\includegraphics[bb={157 465 478 695},scale=1.0]{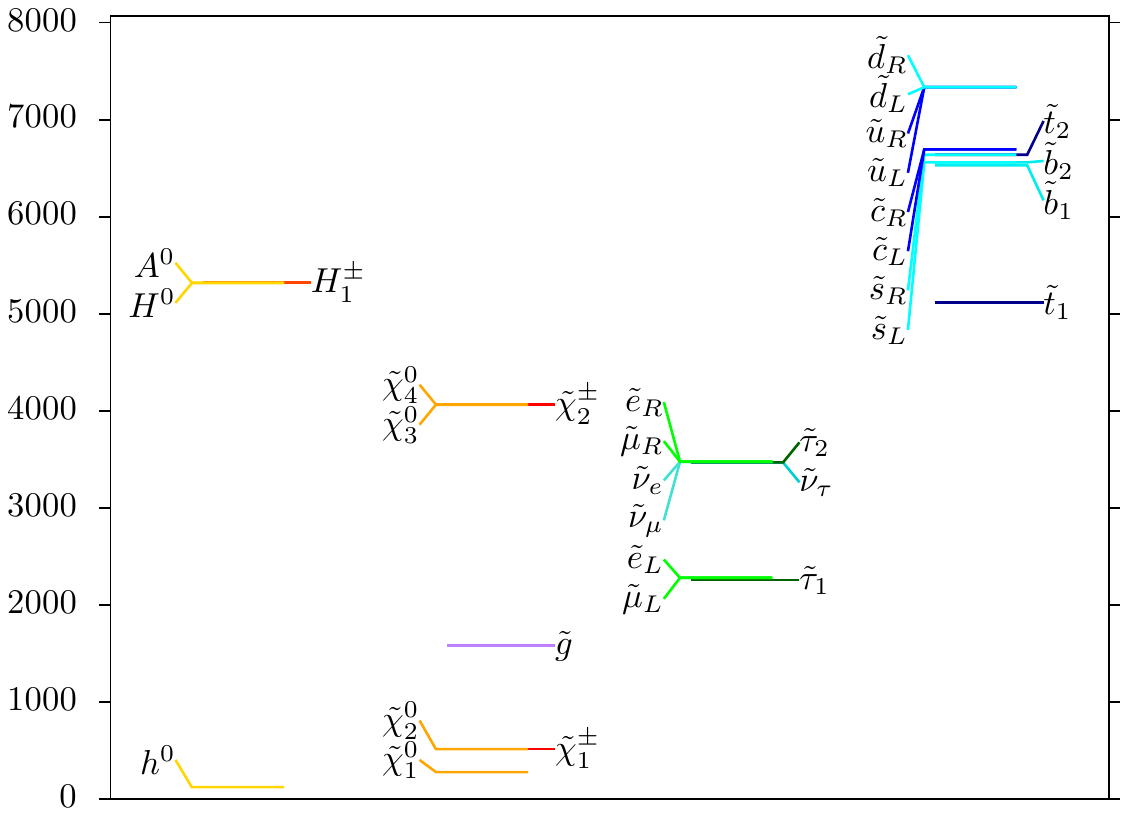}
\end{center}
\caption{Similar to Fig.\ref{fig:spectrum_best}, but for conformal scale $M_{*}=5\times10^{10}$ GeV, in this case the LSP is neutralino. The other parameters are fixed to $\Lambda_{a}=10^6$ GeV, $\epsilon=0.2$ and $\lambda_{u}=0.3$.}
\label{fig:spectrum_high}
\end{figure}

The phenomenology of this model is different from conventional GMSB, since the NLSP (LSP)is neutralino rather than stau in most of the parameter space, which could be seen in Fig.~\ref{fig:spectrum_best}. In the colored sector, the stops are several TeV to satisfy $m_h=125$~GeV. Therefore all the squarks and sleptons escape the current limits of LHC. In Fig.~\ref{fig:spectrum_best}, the spectrum of the model at best point is given, which implies $H^0$, $A^0$ and $H^{\pm}$ are quite heavy so that the Higgs sector is within the decoupling limit and the lightest Higgs properties are those of the Standard Model. In Fig.~\ref{fig:spectrum_high}, the spectra at high conformal scale are given. Here neutralino becomes the LSP, which plays a crucial role in DM research.

\section{Conclusion}
\label{sec:conclusion}
The discovery of the Higgs boson with a mass of $m_h\simeq125$~GeV raises a challenge for naturalness in the MSSM. In order to alleviate the fine tuning induced by several TeV stops, we have introduced a model for focus point SUSY in the context of gauge mediation. In contrast to previous attempts to combine gauge mediation and focus point SUSY our model is very simple but keeps the fine tuning under control.

Through the combination of Yukawa mediation and conformal sequestering, we found a calculable model of gauge mediation, which automatically satisfies the minimal flavor violation (MFV). In addition, the $A$-terms in this model are small for the price of heavy stops. However, this also evades possible issues with color and charge breaking minima. Although the suppression of the $A$-terms, the corresponding fine tuning in this model is signficantly smaller than in minimal GMSB. To demonstrate this we performed a full fledged numerical calculation of the fine tuning in this model using the combination of the public tools {\tt SARAH} and {\tt SPheno}. 

In this paper, we concentrated on a moderate value for the conformal scale, i.e., $M_*<10^{10}$~GeV. There are numerous avenues for exploring models with high conformal scale. In particular, when $M_*>10^{10}$~GeV, the LSP is no longer the Gravitino but the lightest neutralino. Hence, we would have a standard WIMP candidate for dark matter as well.

\section*{Acknowledgements}
TL was supported in part
by the Natural Science Foundation of China
under grant numbers 10821504, 11075194, 11135003, 11275246, 
and by the the National Basic Research Program of China (973 Program) under Grant No. 2010CB833000.
FS is supported by the BMBF
PT DESY Verbundprojekt 05H2013-THEORIE 'Vergleich von LHC-Daten mit supersymmetrischen Modellen'.


\appendix

\section{Conventions and One-Loop RGEs}
\label{sec:convention}
Our convention is the same as that in Refs.~\cite{Kazakov:2000us,Vempati:2000nm}. We define
\begin{align}
t=2\log\left(\frac{M_{*}}{Q}\right),
\label{eqn:runningparameter}
\end{align}
where $M_{*}$ is the conformal scale scale at which the hidden sector renormalization decouples. To simplify the analytical calculation, we only consider the third generation Yukawa couplings and use the notation
\begin{align}
{\mbox a}_i&\equiv
\frac{g_i^2}{16\pi^2}, (i=1,2,3),\\ Y_k&\equiv
 \frac{(y^k_{33})^2}{16\pi^2}, (k=t,b,\tau),\\
\alpha_{\lambda}&\equiv\frac{\lambda_u^2}{16\pi^2},
\end{align}
where $\lambda$ is the marginal coupling between $H_{u}$ and messengers.
The corresponding one-loop RGEs for the MSSM are
\begin{align}
\frac{d{\mbox a}_i}{dt}&=-b_i {\mbox a}_i^2, \nonumber\\ \frac{dY_t}{dt}
& =  Y_{\tau}\left(\frac{16}{3}{\mbox a}_3 + 3{\mbox a}_2 +
\frac{13}{15}{\mbox a}_1-6Y_t-Y_b\right) , \nonumber \\
\frac{dY_b}{dt} & =  Y_b\left(\frac{16}{3}{\mbox a}_3 + 3{\mbox a}_2 +
\frac{7}{15}{\mbox a}_1-Y_t-6Y_b-Y_{\tau}\right), \nonumber
\\ \frac{dY_{\tau}}{dt} & =  Y_{\tau}\left( 3{\mbox a}_2 +
  \frac{9}{5}{\mbox a}_1-3Y_b-4Y_{\tau}\right). \label{eq1}
\end{align}
with
\begin{align}
(b_1,\;b_2,\;b_3)=(33/5,\;1,\;-3)
\end{align}
The $\beta$-functions for the soft terms are given at one-loop by
\allowdisplaybreaks
\begin{align}
\frac{dM_i}{dt} & = - b_i {\mbox a}_iM_i . \nonumber\\
\frac{dA_t}{dt} & =  -\left(\frac{16}{3}{\mbox a}_3 M_3 + 3{\mbox a}_2
M_2 + \frac{13}{15}{\mbox a}_1 M_1+6Y_tA_t+Y_bA_b\right), \nonumber
\\
 \frac{dA_b}{dt} & = -\left( \frac{16}{3}{\mbox a}_3 M_3 +
3{\mbox a}_2 M_2 + \frac{7}{15}{\mbox a}_1 M_1+6Y_bA_b+Y_tA_t+Y_{\tau}A_{\tau}\right),
\nonumber \\ \frac{dA_{\tau}}{dt} & =  -\left( 3{\mbox a}_2 M_2 +
\frac{9}{5}{\mbox a}_1 M_1+3Y_bA_b+4Y_{\tau}A_{\tau}\right), \nonumber
\\ \frac{dB}{dt} & =  -\left(3{\mbox a}_2 M_2 +
\frac{3}{5}{\mbox a}_1 M_1+3Y_tA_t+3Y_bA_b+Y_{\tau}A_{\tau}\right). \nonumber
\\
 \frac{d\tilde{m}^2_Q}{dt} & =
\left(\frac{16}{3}{\mbox a}_3M^2_3 + 3{\mbox a}_2M^2_2 +
\frac{1}{15}{\mbox a}_1M^2_1\right)
-Y_t(\tilde{m}^2_Q+\tilde{m}^2_U+m^2_{H_u}+A^2_t) \nonumber
\\
 &  -Y_b(\tilde{m}^2_Q+\tilde{m}^2_D+m^2_{H_d}+A^2_b), \nonumber\\
\frac{d\tilde{m}^2_U}{dt} & =  \left(\frac{16}{3}{\mbox a}_3M^2_3
+\frac{16}{15}{\mbox a}_1M^2_1\right)
-2Y_t(\tilde{m}^2_Q+\tilde{m}^2_U+m^2_{H_u}+A^2_t) ,
\nonumber\\ \frac{d\tilde{m}^2_D}{dt} & =
 \left(\frac{16}{3}{\mbox a}_3M^2_3+ \frac{4}{15}{\mbox a}_1M^2_1\right)
-2Y_b(\tilde{m}^2_Q+\tilde{m}^2_D+m^2_{H_d}+A^2_b),
\nonumber\\
  \frac{d\tilde{m}^2_L}{dt} & =
3\left(
 {\mbox a}_2M^2_2 + \frac{1}{5}{\mbox a}_1M^2_1\right)
-Y_{\tau}(\tilde{m}^2_L+\tilde{m}^2_E+m^2_{H_d}+A^2_{\tau}),
\nonumber
\\ \frac{d\tilde{m}^2_E}{dt} & =  \left(
 \frac{12}{5}{\mbox a}_1M^2_1\right)-2Y_{\tau}(
\tilde{m}^2_L+\tilde{m}^2_E+m^2_{H_d}+A^2_{\tau}), \nonumber
\\
\frac{d\mu^2}{dt}&=\mu^2\left[3\left({\mbox a}_2+
\frac{1}{5}{\mbox a}_1\right)-(3Y_t+3Y_b+Y_{\tau})\right],\label{eq2} \\
\frac{dm^2_{H_d}}{dt} & =  3\left({\mbox a}_2M^2_2
+\frac{1}{5}{\mbox a}_1M^2_1\right)
-3Y_b(\tilde{m}^2_Q+\tilde{m}^2_D+m^2_{H_d}+A^2_b)
\nonumber
\\ &- Y_{\tau}(\tilde{m}^2_L+\tilde{m}^2_E+m^2_{H_d}+A^2_{\tau}) ,
\nonumber\\ \frac{dm^2_{H_u}}{dt} & =  3\left({\mbox a}_2M^2_2
+\frac{1}{5}{\mbox a}_1M^2_1\right)
-3Y_t(\tilde{m}^2_Q+\tilde{m}^2_U+m^2_{H_u}+A^2_t),
\nonumber
\end{align}
\section{General Derivation of Focus Point Formula}
\label{sec:derivation}
In this appendix, we reproduce the well-known formula for focus point SUSY proposed in Refs.~\cite{Feng:1999mn,Feng:1999zg,Feng:2000bp} in the context of our conventions of Sec.~\ref{sec:convention}. In the region of small $\tan\beta$, the RGEs can be solved analytically \cite{Ibanez:1984vq}
\begin{align}
{\mbox a}_i[t]& = \frac{{\mbox a}_i[0]}{1+{\mbox a}_i[0]b_it},\\
Y_t[t]&=\frac{Y_{t}[0] E[t]}{1+6Y_t[0]F_[t]},
 \label{sol}
\end{align}
where
\begin{align}
E[t]&=\prod_i(1+b_i{\mbox a}_i[0]t)^{c_i/b_i} ,\\
c_{i}&=\left(\frac{13}{15},3,\frac{16}{3}\right), \\
F[t]&=\int^t_0 E[t']dt'.
\end{align}
Therefore, we have
\begin{align}
\mathcal{I}&=\exp\left(-6 \int_0^t Y_t[t']dt'\right)\nonumber\\
           &=\exp\left(-6 \int_0^t \frac{Y_t[0]E[t']}{1+6 Y_t[0]F[t'] }dt' \right)\nonumber\\
           &=\frac{1}{1+6 Y_t[0]F[t]}\nonumber\\
           &=1-\frac{6Y_t[t]F[t]}{E[t]}\label{eqn:focuspoint}
\end{align}
The formula Eq.~(\ref{eqn:focuspoint}) plays a crucial role in determining whether or not focus point supersymmetry is available in a given model. Note, compared to Ref.~\cite{Feng:1999zg} there is an opposite sign. The reason is the different definition of running parameter in Eq.~(\ref{eqn:runningparameter}).

%

\bibliography{FP_SUSY}
\bibliographystyle{ArXiv}

\end{document}